\documentclass[12pt]{article}
\usepackage{graphicx}
\begin{document}
\begin{center}

{\large \bf Resonances in Ps-H scattering}  

\vskip 0.3cm

{\bf Hasi Ray}\\

Institute of Plasma Research, Bhat, Gandhinagar, Ahmedabad 382428, INDIA 

{\bf Email: hasi\_ray@yahoo.com  \& hasiray@ipr.res.in}
\end{center}
\vskip 0.25cm
{\bf Abstract:
	The best 3-channel projectile-inelastic [H(1s)+Ps(1s,2s,2p)] close-coupling approximation (CCA) is performed for positronium (Ps) and hydrogen (H) collision considering both the atoms in ground states at the incident channel. The s-wave elastic phase shifts and cross sections in the singlet and triplet channels and the total (or integrated) cross section are studied below inelastic threshold. Resonances in singlet channel using 2-channel and 3-channel projectile-inelastic CCAs confirm earlier prediction [1]. A resonance in triplet channel using 3-channel projectile-inelastic CCA is a new addition.
} \\


\vskip 0.5cm
\section{Introduction}
To learn new physics, the Ps-H is an ideal system since the target(H) and the projectile(Ps) both are hydrogen-like systems with a single electron; their wavefunctions are known exactly. Being a bound system of a particle(electron) and its antiparticle(positron), the charge and mass centers coincide in positronium(Ps); this gives rise to a zero mean static potential. In addition the eight times higher polarizability than hydrogen(H) makes Ps an interesting probe for new investigations. The theoretical studies of such a system is rigorous due to the presence of four charge centers. At very low energies near to zero although elastic scattering is the only active channel but the excitation and ionization thresholds are very close. So the quantum mechanical influence of different inelastic channels are important and should be taken into consideration. It is a challenging problem to get accurate low energy data for such a system.

In collision physics, resonance is an important phenomenon. When a microscopic moving object that is a wave, enters into the scattering chamber near the target, it faces interactions. When it comes out of the scattering zone, the original incident wave gathers a phase shift and the new wave is known as scattered wave. The change in phase which is named as phase shift, is the parameter that carries the information of the scattering process. A rapid change in phase shift by $\pi$ radian in a very narrow energy interval of the incident wave is an indication of the existence of a resonance. It carries the information of a bound system if in the s-wave elastic scattering and below threshold of excitation. One can calculate the width of resonance to get the life time ($\tau$) of the newly formed system.

It is an extremely difficult job to detect a resonance since successful identification needs (i) a very accurate calculation and (ii) sufficient computation facilities. A large number of mess-points in a very small energy interval, generally $\sim 10^{-2} - 10^{-3}$ eV is required. It necessitates a high-speed computer with a sufficient memory, the knowledge of mathematical computation and programming languages e.g. FORTRAN. 

We are interested to discuss the scattering processes at low energies. At high energies the projectile usually ignores all the important delicate interactions with target, so generally it carries no valuable information. At very low energies below excitation threshold, elastic scattering is the only real process. But from just above the threshold different excitation, ionization etc. channels start to open. It is a very difficult task to study the scattering processes at intermediate energies due to presence of many different channels and a very close coupling among them. The total cross section is the sum of the integrated cross sections of all these channels. Besides these, the partial wave contribution from higher angular momenta start to dominate since above the threshold. Many close peaks may arise in the cross section at the intermediate energies due to opening of different channels and their mixing. These are also named as resonances and studied by Higgins \& Burke [2-3], Sarkar et al [4-5]. However Zhou et al [6] commented against such resonances. So the subject of above threshold resonances needs more investigation. Above threshold, the phase shift becomes a complex quantity and known as eigen phase shift.
\subsection{Resonances}
The kind of resonances we are interested in, is feasible only below inelastic threshold. At this energy region, the integrated elastic cross section ($\sigma_{el} = \sigma_{L=0}^{el} + \sigma_{L=1}^{el} + \sigma_{L=2}^{el} + ... $) is the total cross section ($\sigma_{Total} = \sigma_{el} + \sum\sigma_{inelastic}$) due to presence of only elastic channel; again s-wave (i.e. $L=0$ partial wave) dominates. So effectively, the projectile angular momentum $L = 0$ partial wave carries almost all the information of the scattering process.

The phase shift $\delta_l$ can be decomposed as 
 
$\delta_l = \xi_l + \eta_l $.\\
$\xi_l$ corresponds to the hard sphere scattering or non-resonant part; it does not depend on the shape and depth of the potential. The term $\eta_l$ depends on the details of the potential. The quantities $\xi_l$ and $\eta_l$ vary in general slowly and smoothly with the incident particle energy. But in certain cases $\eta_l$ may vary rapidly in a small energy interval of width $\Gamma$ about a given energy value $E_R$ such that we can write
$$ \eta_l = \eta_l^R = tan^{-1}\frac{\Gamma}{2(E_R-E)} $$\\
In that energy interval the phase shift is therefore given approximately by\\ 
$\delta_l \simeq \xi_l+\eta_l^R$.

 The physical significance of a narrow resonance can be inferred by examining the amplitude of the radial wave function inside the interaction region. The probability of finding the scattered particle within the potential is much larger near the resonance energy $E=E_R$, so that in that case the particle is nearly bound in the well. Thus the resonance may be considered as a metastable state whose lifetime $\tau$, which is much longer than a typical collision time, can be related to the resonance width $\Gamma$ by using the uncertainty relation ${\Delta}t{\Delta}E\geq\hbar$. Thus, with ${\Delta}t\simeq\tau$ and ${\Delta}E\simeq\Gamma$, we have $\tau\simeq\frac{\hbar}{\Gamma}$.

The shape of the cross section curve near a resonance as a function of energy depends on the non-resonant phase shift $\xi_l$ [7,8]. For the s-wave scattering it is 
$$ \sigma_l = \frac{sin^2\xi_l(E_R-E)^2 + cos^2\xi_l\Gamma^2 + sin2\xi_l(E_R-E)\Gamma/2}{(E_R - E)^2 + \frac{\Gamma^2}{4}}$$

Two limiting cases for non-resonant phase shift are 0 and $\pi/2$. In the first case the above equation becomes 

         $$\sigma_l = \frac{\Gamma^2}{(E_R - E)^2 + \frac{\Gamma^2}{4}}$$\\ 
which is symmetric and represents a rise in cross section at the resonance energy. In the other case 

         $$\sigma_l = \frac{(E_R - E)^2}{(E_R - E)^2 + \frac{\Gamma^2}{4}}$$\\ 
which is also symmetric but goes down to zero at the resonance energy. If the non-resonant phase shift gets some other value then all sorts of forms of the cross section can occur.
 
 However one needs to investigate both the phase shift and the cross section, but should be careful to use the formulation which derives the cross section directly from amplitude and not from phase shift. If resonance is true it should be reflected both in phase shift and in cross section. It is actually to make a precautionery measure for successful detection since the calculation of phase shift involves a tan-inverse function which may create a numerical error. 

\subsection{Close coupling approximation (CCA)}
  The CCA is a highly successful theory to study low energy scattering phenomenon in atomic physics. The theory was predicted by Massey [9] and applied in $e^-$ - atom scattering. Later Burke et al [10] successfully used this theory for $e^+$ - atom scattering. Now a days many different groups of the world are using this theory for $e^-$/$e^+$ - atom scattering. It was Fraser [11] who used it first for Ps - H scattering. Ray \& Ghosh [12-13] merits the credit because they are the first who supplied detailed  and converged results. They used a momentum space formalism introduced by Calcutta group [14] whereas Fraser used a coordinate space formalism to write the coupled integral equations. They again add more channels in the CCA basis [15-28]. The studies of Fraser [11,29-30], Fraser et al [31], Hara et al [32] were confined to static-exchange model i.e. considering only the elastic channel in the basis and the H and He targets. Ray [16-18] extended the CCA theory in Ps and lithium (Li) scattering using the static-exchange and a two-channel CCA models.  

  The theory is based on the very basic principle of quantum mechanics i.e. the eigen state expansion (ESE) methodology in which the total wave function of a quantum mechanical system is expressed as a linear combination of all possible states known as basis set. So one has to use a wide channel space, but practically it is impossible. The number of unknowns exceed the number of equations when non-spherical orbitals like p- and d- states of an atom are considered in channel space. So arised the necessity of an approximation. We should conserve the total angular momentum quantum numbers e.g. `J` and `M`. It makes the equations closed i.e. the number of unknowns are equal to the number of equations. The accuracy of the method depends on the choice of basis set. It should be remembered that when we study a particular channel (e.g. suppose elastic channel), the quantum mechanical effect of other included channels (through basis set) comes to play due to angular momentum coupling through conservation of good quantum numbers 'J' and 'M'.

Two interacting atoms mutually induce a symmetrical pair of dipoles [33]. The induced polarization potential is the same for electron-atom and positron-atom scattering, in both cases they are attractive [34]. Ps is a highly polarizable atom. In coupled channel methodology, the inclusion of the effect of dipole polarizability of an atom is possible through excitation of the atom to different p-orbitals[10,15-18,35-37]. The inclusion of different excited s-states provides a short-range strong-coupling force [10] whereas the excited p-states supply both a short-range as well as a long-range forces [10,15-17,26-27]. The mean static interaction provides a null contribution if Ps. Earlier [15] we investigated Ps-H scattering using target-inelastic CCAs to study the effect of H(2s) and H(2p) states on elastic channel using target-inelastic CCAs. Investigations performed by Burke et al [10,35-36] on $e^--H$ and $e^+-H$ systems established that the inclusion of only the first excited p-state of H i.e. H(2p), in the CCA scheme can include $65.77\%$ effects of dipole polarizability. Similar properties are expected from Ps as an isotope of H. To include the total effect of dipole polarizability one has to choose a CCA basis with an infinite number of p-states of both the atoms; it is definitely impossible. So one can increase the accuracy of the calculation by expanding the basis set only. However, according to the concept of quantum mechanical approach, the next target-elastic Ps(3p) channel is most useful. The first order effect due to static dipole polarizability vanishes in Ps-atom systems. On the otherhand CCA is highly efficient to include the static as well as dynamic or non-adiabatic effects [36,38]. Non-adiabatic effect is important [17,38-40] if highly polarizabable atom (e.g. Ps). 

The electrons are indistinguishable particles. Two electrons can interchange their positions, the phenomenon is known as exchange. At lower energies the effect of exchange is highly important when more than one electron in the system. The spins of the target and the projectile electrons can make a distribution of 1/4 possibility of forming a singlet (+) state (combined spin = 0) and 3/4 possibility of forming a triplet (-) state (combined spin = 1). Accordingly the space part of the system wave function should take a symmetric and an antisymmetric form to make the total wave function of the system an antisymmetric one.

 We consider a 3-channel projectile-inelastic basis as [H(1s)+Ps(1s,2s,2p)] with all the nine possible primary 

$$
\begin{array}{cc}
     Ps(1s) + H(1s) & \rightarrow Ps(1s) + H(1s)\\
     Ps(1s) + H(1s) & \rightarrow Ps(2s) + H(1s)\\
     Ps(1s) + H(1s) & \rightarrow Ps(2p) + H(1s)\\
\end{array}\\
$$
and intermediate channels 

$$
\begin{array}{cc}
     Ps(2s) + H(1s) & \rightarrow Ps(1s) + H(1s)\\
                    & \rightarrow Ps(2s) + H(1s)\\
                    & \rightarrow Ps(2p) + H(1s)\\
     Ps(2p) + H(1s) & \rightarrow Ps(1s) + H(1s)\\
                    & \rightarrow Ps(2s) + H(1s)\\
                    & \rightarrow Ps(2p) + H(1s) 
\end{array} $$ 
\\
channels. It is the best 3-channel basis according to concept of quantum mechanical approach of eigen-state expansion. All the possible Coulomb interactions are treated exactly for all the direct and exchange matrix elements. Earlier study [26] using such a methodology neglected the most important intermediate channel [Ps(2p)+H(1s)$\rightarrow$Ps(2p)+H(1s)] which is responsible for non-adiabatic or dynamic effects [38-40]. 

\section{Theory}
The total wavefunction of the system $\Psi^{\pm}$ 
satisfying the Schr{$\ddot o$}dinger equation:
\begin{equation}
H{\Psi}^{\pm}({\bf r_p},{\bf r_1},{\bf r_2}) = E
{\Psi}^{\pm}({\bf r_p},
{\bf r_1},{\bf r_2})
\end{equation}
is expressed as 
\begin{eqnarray}
\Psi^{\pm}({\bf r_p,r_1,r_2}) 
=\frac{1}{\sqrt 2}(1{\pm}P_{12})\sum_{n_t l_t n_p l_p L J_1 J M} \frac{F_{\Gamma_0 n_t l_t n_p l_p L J_1 J M}({\bf k},{\bf k^\prime},R_1)}{R_1}
\frac{U_{n_t l_t}(r_2)}{r_2} \frac{V_{n_p l_p}(\rho_1)}{\rho_1} \nonumber \\
\sum_{m_t m_p M_L}
\left(
\begin{array}{ccc}
  L & l_p & J_1 \\
  M & m_p & M_1 
\end{array}\right)
\left(
\begin{array}{ccc}
  J_1 & l_t & J \\
  M_1 & m_t & M 
\end{array}\right)
 Y_{LM_L}({\hat{\bf R}_1}) Y_{l_p m_p}(\hat{\mbox\boldmath\rho_1})Y_{l_t m_t}(\hat{\bf r}_2)
\end{eqnarray}
with
\begin{equation}
H = -\frac{1}{2}\mbox{\boldmath $\nabla_p^2$}
-\frac{1}{2}\mbox{\boldmath $\nabla_1^2$}
-\frac{1}{2}\mbox{\boldmath $\nabla_2^2$}+\frac{1}{\mid{\bf r_p}\mid}
-\frac{1}{\mid{\bf r_1}\mid}-\frac{1}{\mid{\bf
r_2}\mid}
-\frac{1}{\mid{\bf r_p-\bf r_1}\mid}-\frac{1}{\mid{\bf
r_p-\bf r_2}\mid}
+\frac{1}{\mid{\bf r_1-\bf r_2}\mid} 
\end{equation}
Here $P_{12}$ stands for the exchange operator and
${\bf R_i}=\frac{1}{2}({\bf r_p}+{\bf r_i})$ and
$\mbox{\boldmath $\rho_i$}={{\bf r_p}-{\bf r_i}}$; i=1,2. ${\bf r_1}$ and
${\bf r_2}$ are the
position vectors of the electrons belonging to Ps and
H respectively and ${\bf r_p}$, is that of
the positron with
respect to the center of mass of the system.
${U_{n_t l_t}({\bf r})/r}$ and $V_{n_p l_p}(\mbox{\boldmath$\rho$})/\rho$
are the radial parts of the wavefunctions
of H and Ps respectively and $F_{\Gamma_0 \Gamma}({\bf k,k^\prime},R)/R$
is the radial part of the continuum wave
function of the moving Ps atom; $\Gamma_0$ indicates all the quantities $n_t l_t n_p l_p L J_1 J M$ of $\Gamma$ at the initial channel.
 
Projecting the Schr$\ddot o$dinger Eqn.(1) just like the Hartree-Fock variational approach and integrating over the desired coordinates, we can get a set of integro-differential equations. These integro-differential equations can be transformed into the integral equations applying the asymptotic boundary conditions like Lippmann-Schwinger [14]. These coupled integral equations can be formed either in momentum space or in configuration space. Fraser et al used the configuration space approach whereas we are using momentum space formalism [14]. The set of coupled integral equations obtained for the scattering amplitudes are as follows:
\begin{equation}
f^{\pm}_{n^{\prime}l^{\prime},nl}({\bf k^{\prime}},{\bf k})
=B^{\pm}_{n^{\prime}l^{\prime},nl}({\bf k^{\prime}},{\bf
k})
-{\frac{1}{2\pi^2}}\sum_{n''}{\int}d{\bf
{k''}}\frac{B^{\pm}_{n'l',n''l''}({\bf k'},{\bf
k''})f^{\pm}_{n''l'',nl}({\bf k''},{\bf
k})}{k^2_{n''l''}-{k''}^2+i{\epsilon}}
\end{equation}
$B^{\pm}$ indicate the Born-Oppenheimer [41] scattering amplitudes; plus(+) is for the singlet channel for which the space-part of the system wave function is symmetric, and minus(-) is for triplet channel for which the space-part of the system wave function is antisymmetric. The formulation for Born matrix element is available in our previous articles [42-46] and Born-Oppenheimer in recent articles [47,48]. Similarly $f^{\pm}$ indicate the unknown scattering amplitudes for the singlet and the triplet channels respectively. The summation over $n''l''$ is to include various channels.

These three dimensional coupled integral equations involving $f_{n^{\prime}l^{\prime},nl}({\bf k^{\prime}},{\bf k})$ and $B_{n^{\prime}l^{\prime},nl}({\bf k^{\prime}},{\bf k})$ can be reduced to the corresponding one dimensional forms through partial wave analysis using the expansion like:  
                                                                  \begin{eqnarray}                                                                   \tau^{\pm}_{n^{\prime}l^{\prime}m^{\prime},nlm}({\bf k^{\prime}},{\bf k})=(kk^{\prime})^{-1/2}\sum_{J,M}\sum_{L,M_L}\sum_{L^{\prime},M_L^{\prime}}\sum_{J_1,M_1}\sum_{J_1^{\prime},M_1^{\prime}} \langle L^{\prime}l_p^{\prime},M_L^{\prime}m_p^{\prime}|J_1^{\prime}M_1^{\prime}\rangle \langle Ll_p,M_Lm_p|J_1M_1\rangle \nonumber \\     \langle J_1^{\prime}l_t^{\prime},M_1^{\prime}m_t^{\prime}|JM\rangle \langle J_1l_t,M_1m_t|JM\rangle Y_{L^{\prime}M_{L^{\prime}}}( \hat{\bf k^{\prime}}) Y_{LM_L}( \hat{\bf k}) \tau^{J\pm}(k^{\prime}n_p^{\prime}n_t^{\prime}l_p^{\prime}l_t^{\prime},kn_pn_tl_pl_t)                                              \end{eqnarray} 
 Here ${\bf\left[A\right]}$ is the scattering matrix of $N{\times}N$ type formed by Born and Born-Oppenheimer [41] scattering amplitudes, ${\bf \left[X\right]}$ is the column matrix of $N\times1$ type formed by unknown CCA scattering amplitudes and ${\bf \left[B\right]}$ is again a column matrix of $N\times1$ type formed by the Born and Born-Oppenheimer [41] scattering amplitudes; the dimension(N) depends on the number of channels included in the expansion basis. We have calculated all the Born and Born-Oppenheimer amplitudes exactly following an analytic approach and then the rest part is carried on following numerical approach using computer and FORTRAN programming.
 The two sets of one dimensional coupled integral equations of scattering amplitudes in momentum space for the singlet(+) and triplet(-) channels respectively, are solved separately for each partial wave(L).  
  
\section{Results and discussion}
\begin{figure*}
\centering
\includegraphics[width=0.95\columnwidth]{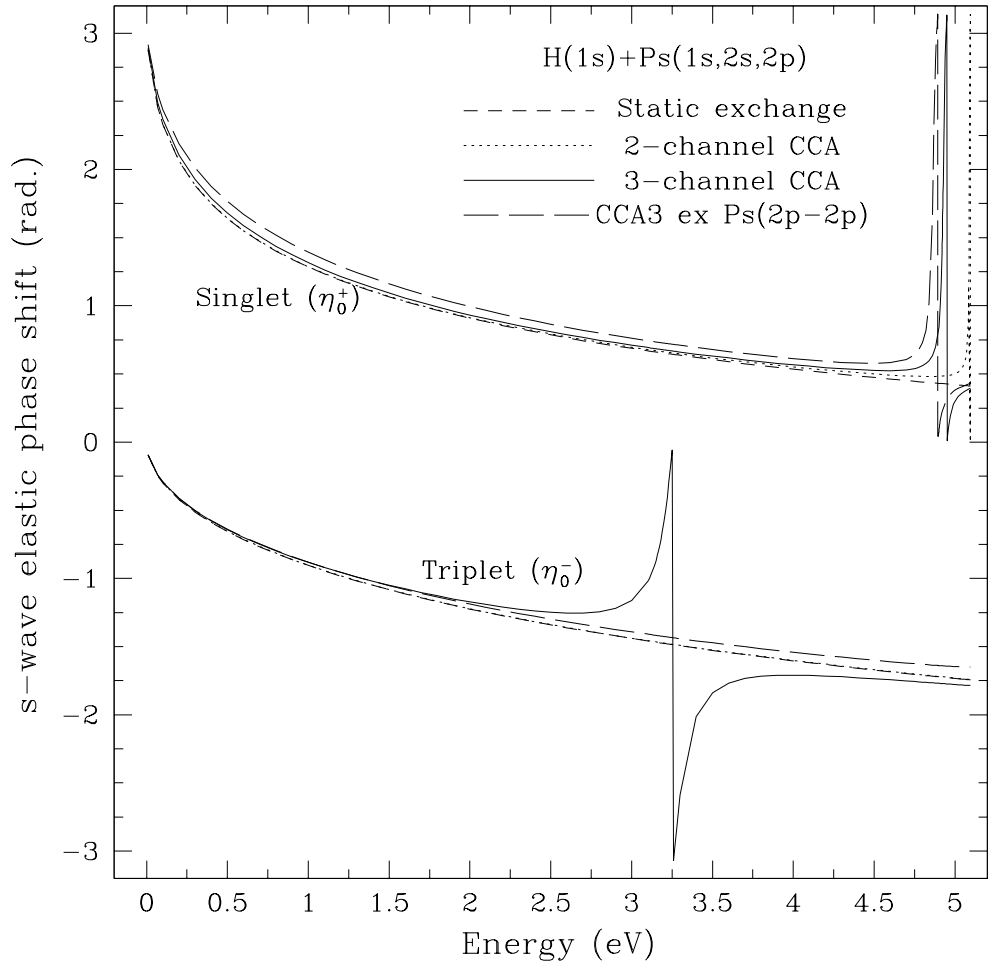}
\caption{
 The s-wave elastic phase shifts for Ps-H scattering below threshold.}%
\end{figure*}

\begin{figure*}
\centering
\includegraphics[width=0.95\columnwidth]{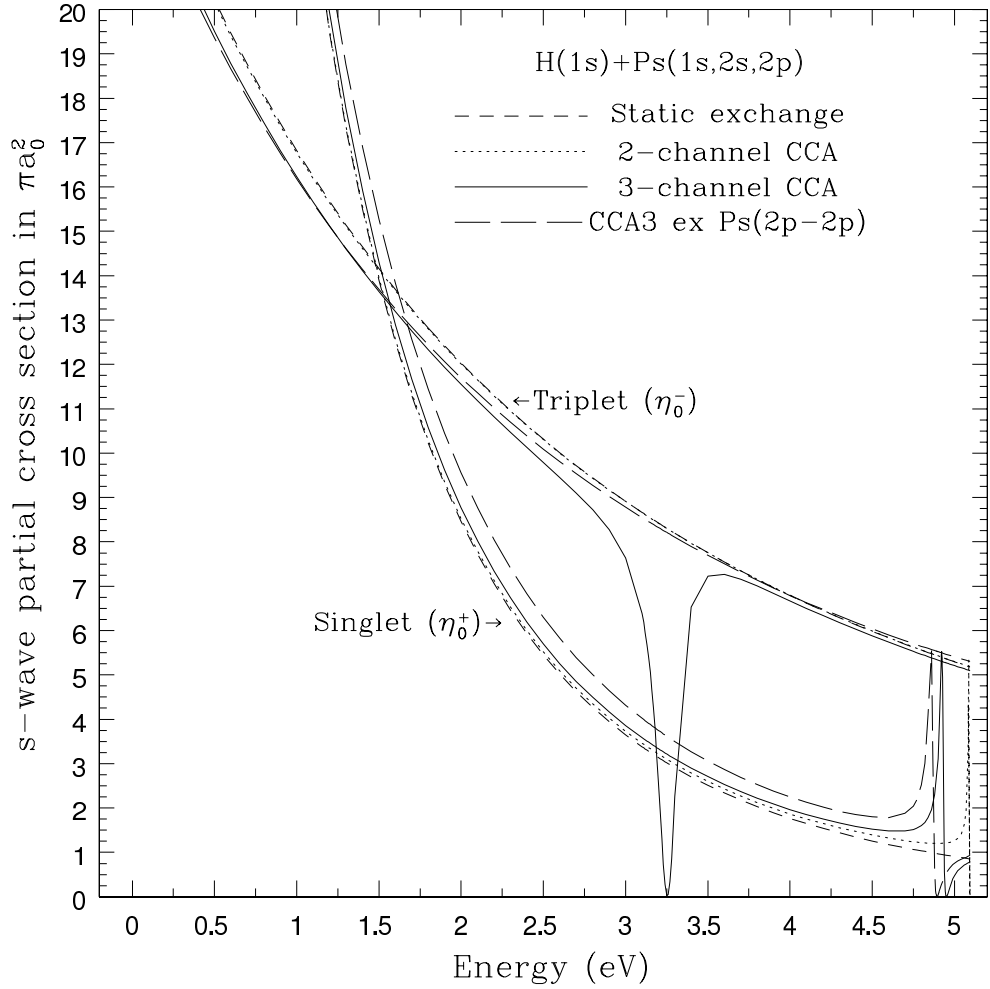}
\caption{
 The s-wave elastic cross sections for Ps-H scattering below threshold.}
\end{figure*}

\begin{figure*}
\centering
\includegraphics[width=0.95\columnwidth]{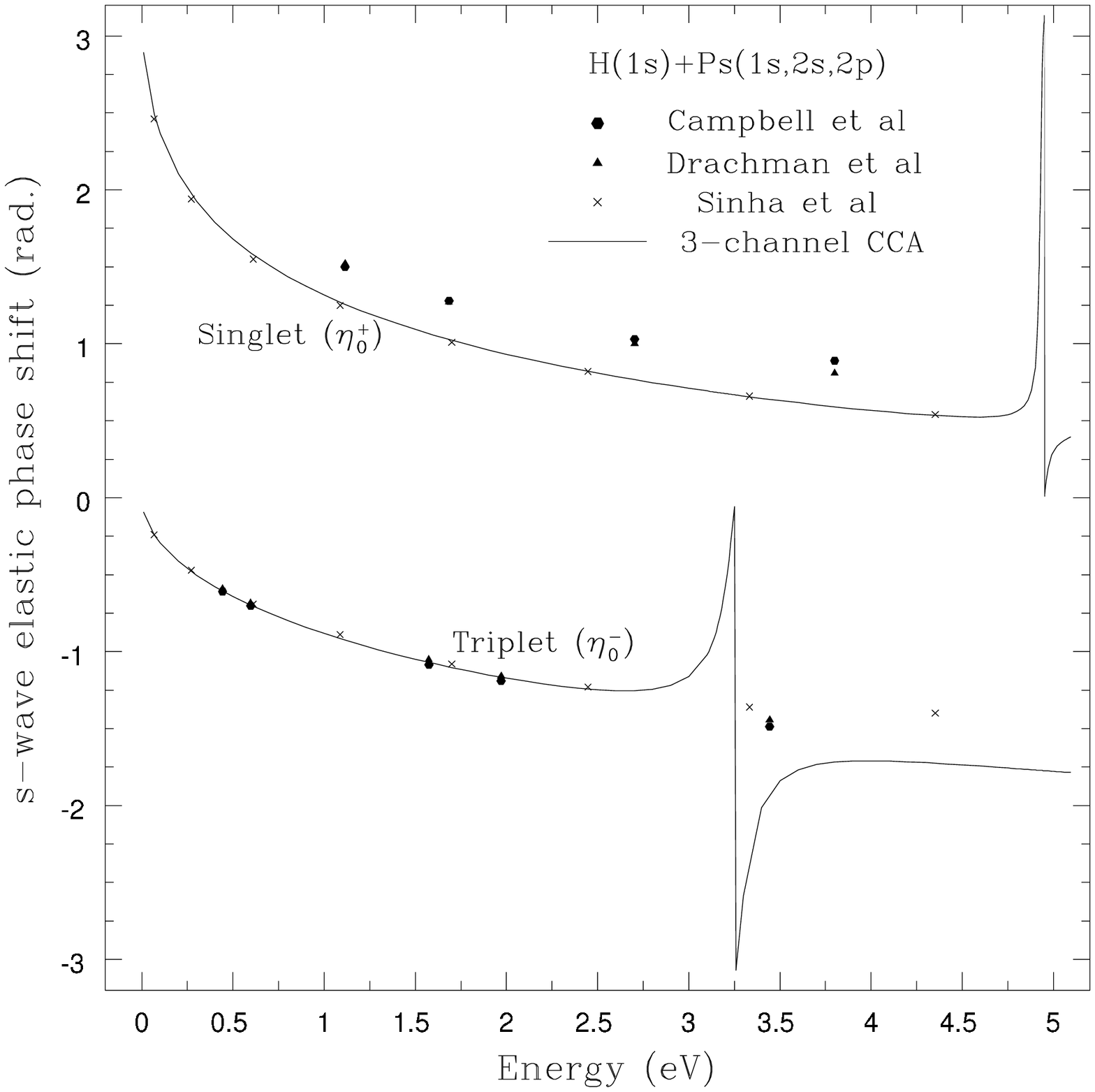}
\caption{
 The comparison of s-wave elastic phase shifts for Ps-H scattering below threshold.}
\end{figure*}

\begin{figure*}
\centering
\includegraphics[width=0.95\columnwidth]{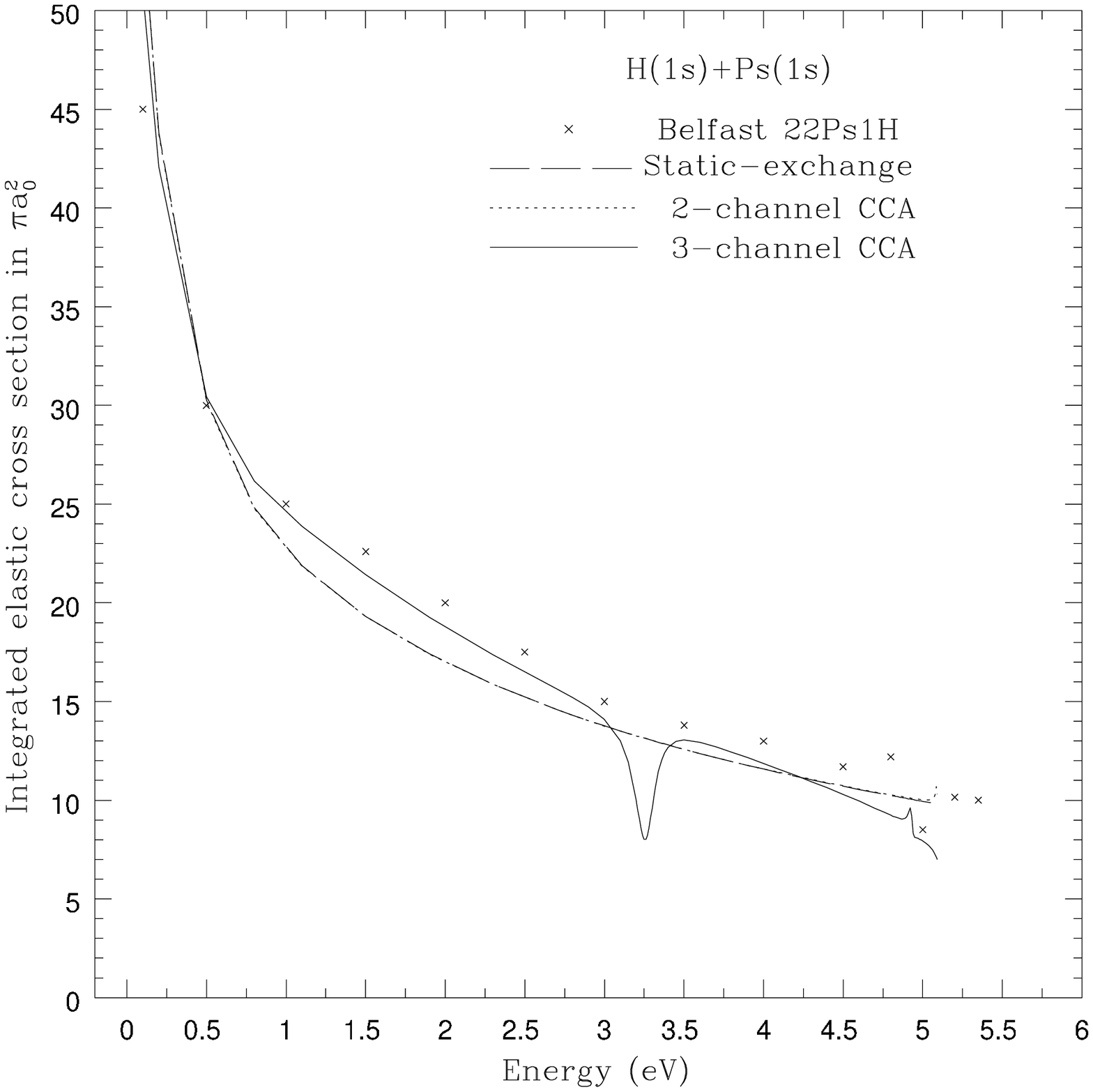}
\caption{
 The total cross sections for Ps-H scattering below inelastic threshold.}
\end{figure*}

We find exact resonances both in singlet and triplet channels in Ps-H scattering below inelastic threshold in s-wave elastic phase shifts as well as in cross sections. The singlet resonances add more importance on the predicted binding [1]. The triplet resonance predicts a new binding. Attached Figure 1 shows our s-wave elastic phase shifts below inelastic threshold and Figure 2, the corresponding s-wave partial cross sections. 
The phase shifts and cross sections, both are derived directly from the elastic scattering amplitudes. Both the singlet(+) and triplet(-) results are presented using the static-exchange approximation, the 2-channel projectile-inelastic CCA (CCA2), the 3-channel projectile-inelastic CCA (CCA3) and using CCA3 but excluding the 9th intermediate $Ps(2p)+H(1s)\rightarrow Ps(2p)+H(1s)$. In the first method, only the elastic channel is considered in the CCA basis. In the second method, the first two primary channels are taken with the intermediates and in the third method, all the nine primary and intermediate channels come to play.

In figure 3, both the singlet and triplet phase shifts of the present CCA3 are compared with other theoretical investigations of Campbell et al [49], Drachman et al [1,50] and Sinha et al [26].
The singlet results using CCA2 and CCA3 and the triplet results using CCA3 are in complete agreement with earlier investigation[26] at lower incident energies near to zero but differing near and above the resonance energy region. No phase shift data are available at this energy region near to threshold for the singlet channel [26,49,51]. The triplet result is showing good agreement with other theories but differing only near and above the resonance energy region. 

 In figure 4, the integrated cross sections using present CCAs are compared with the static-exchange and the 22Ps1H coupled-pseudostate R-matrix results of Belfast group. 
Our integrated/total cross sections are also in good agreement with the 22Ps1H coupled-pseudostate R-matrix results [51] below the energy region of triplet resonance. The deviation of the present features at the energy region near and above the triplet resonance welcomes a special attention.

 The existence of a bound system of Ps and H formed by the electrostatic attraction of $e^+$ and $H^-(^1S_0)$ in ground (singlet) state was earlier predicted by Drachman [1] and confirmed by others [49-58]. It is first time observed using a CCA theory. The present study assures that the cause of this singlet resonance is the short-range strong-coupling force arising due to the inclusion of Ps(2s) channel with exact exchange and in complete agreement with earlier prediction. The addition of Ps(2p) channel with exact exchange shifted the position of resonance and broadened the width; these findings are also consistent with the existing physics. The partial cross section is satisfying a formulation like Breit-Wigner which is consistent with a value of non-resonant phase shift of $\sim$ 0.5 radian near the resonance energy. The effect of this resonance is also reflected on our total/integrated cross sections in figure 4. The kink in the total cross section curve near the singlet resonance was also observed by Sinha el al [26] in their 2-channel CCA calculation which is present in our both the CCA2 and CCA3 total cross sections in figure 4. The present findings of singlet resonance in quite agreement with the reality ascertain the accurateness of our present calculation and the computer code.

    The resonance in the triplet(-) channel occur in the CCA3 method. It signifies clearly the cause of this resonance as the inclusion of Ps(2p) channel in the CCA scheme. The inclusion of this channel can provide almost 2/3 effect[10,35-37] of high polarizability of Ps in a dynamic way [36]. If we exclude the intermediate channel involving target-elastic $Ps(2p)\rightarrow Ps(2p)$ transition, the triplet resonance disappears which are displayed in figures 1 \& 2 as 'CCA3 ex Ps(2p-2p)'.  The s-wave triplet cross section at the resonance energy region follows a similar resonance rule for cross section like Breit-Wigner and consistent to the value of non-resonant phase shift $\sim -\pi/2$. The effect of this resonance is again reflected in total/integrated cross sections in figure 4 as a pronounced dip or a minimum in curve. The importance of non-adiabatic effects in the scattering by a highly polarizable target is discussed in Refs.[38-40]. It was established on comparing the results obtained from the non-adiabatic theory by Walters [38] and the CCA theory by Burke and Taylor [36] that the CCA theory is quite capable of including the dynamical effects by its own nature.

In addition, the present triplet phase shift data for Ps-H scattering using the present 3-channel projectile-inelastic CCA fit nicely [59] with non-resonant part as                 $$ \xi_0 = - 1.4053 + 0.1295 E - 0.04613 E^2 $$ \\                  
 and provides the width $\Gamma = 0.15173 eV $ and resonance position $E_R = 3.2630 eV$.

A discussion on various existing theories should be useful. A coupled state representation of the system wavefunction including the effect of important channels is most useful for reliable information. R-matrix theory used by Belfast group [49,51] with a coupled pseudostate formalism of the system wavefunction is highly useful in this respect. In this theory, the system wavefunction is treated accurately upto a certain radius($r\le a$) and above this radius($r > a$) the system wavefunction is treated in an approximate way. But the effect of the long-range interactions are taken into consideration as potentials into the Hamiltonian. The CCA theory used by Calcutta group [12-13,15-18,26-27] and the Sao Paulo group [28] is an approach where the full configuration space is used through the momentum space formalism [14] to get proper long-range effect. The difference should be noted between the long-range effect and the long-range interactions; the entire configuration space should be used for long-range effect with exact interaction but one can approximate the effect of long-range interactions following a trick. The convergence problem due to presence of sine and cosine terms in the configuration space approach to solve the coupled equations at higher angular momentum is discussed by Fraser [11,29-30] and Fraser et al[31-32]. The convergence is easily obtainable using a momentum space formalism. The Sao Paulo group [28] has considered the effect of exchange in an approximate way using a model potential. The methodology used by Calcutta group seems to be more appropriate; we adapt this methodology. 
   The complex rotation and stabilization methods followed by NASA group[1,49,53-55] and Taiwan group[56-57] using Feshbach formalism is a highly efficient tool to detect a Feshbach resonance. Other groups of the world e.g. Australia group[60-63], UCL theory group[64], Italy group[65] are using different variational methods. 

An interesting question is the reality of these resonances i.e. if we increase the basis set whether the resonances still exist. In atom-atom scattering van der Waals interaction is important which comes into picture through excitation of both the atoms to excited p-states. In principle, the major quantum mechanical effect of channel-coupling comes from the closest excitation channels i.e. H(1s)+Ps(1s)$\rightarrow$H(1s)+Ps(2s) and \\
H(1s)+Ps(1s)$\rightarrow$H(1s)+Ps(2p). The target excitation channel and the excitation of both the atoms channel which is responsible for van der Waals interaction are energetically far away from the elastic channel H(1s)+Ps(1s)$\rightarrow$H(1s)+Ps(1s). So the effect of these target excitation and both excitation channels can not be more important in Ps-H scattering. In addition, the reality of singlet resonances are certain. 
\section{Conclusion}
To conclude, we introduce a complete treatment of the best 3-channel projectile-inelastic CCA for Ps-H scattering considering both the atoms in ground states at the initial channel. The scattering phenomena below inelastic threshold have been studied. We see resonances in both the singlet and triplet channels in the s-wave elastic phase shift as well as in s-wave elastic cross section. Both the resonances are again reflected on the total cross section as a peak and a well. Our singlet resonances support the previously predicted singlet resonance [1]. The very interesting newly found triplet resonance predicts a binding; the cause appears to be a non-adiabatic/dynamical force. More investigations using larger basis sets are useful to better understand the fact and the mechanism. The minima in s-wave partial cross section for triplet channel and in integrated/total cross section may indicate a condensation/clustering of the system constitutents. Our theory should be applicable to slowly moving atoms and a dilute system in which the inter atomic separation is much greater than the atomic dimension.
\\

{\bf Acknowledgement} 

Author acknowledges the financial support from DST, India through Grant No. SR/ FTP/ PS-80/ 2001. She is thankful to R. J. Drachman for valuable discussions.
\\

{\bf References:}

1. R. J. Drachman and S. K. Houston, Phys. Rev. A, {\bf 12}, 885 (1975).

2. K. Higgins and P. G. Burke, J. Phys. B {\bf 24}, L343 (1991).

3. K. Higgins and P. G. Burke, J. Phys. B {\bf 26}, 4269 (1993).

4. N. K. Sarkar, Madhumita Basu and A. S. Ghosh, J. Phys. B, {\bf 26}, L427 (1993).

5. N. K. Sarkar, Madhumita Basu and A. S. Ghosh, J. Phys. B, {\bf 26}, L799 (1993).

6. Yan Zhou and C. D. Lin, J. Phys. B, {\bf 28}, L519 (1995).

7. R. J. Drachman, private communication (2005).

8. H. Ray, Phys. Rev. A {\bf 73}, 064501 (2006).

9. H. S. W. Massey and C. B. O. Mohr, Proc. Roy. Soc., {\bf 136}, 289 (1932).

10. P. G. Burke and K. Smith, Rev. Mod. Phys., {\bf 34}, 458 (1962).

11. P. A. Fraser, Proc. Phys. Soc., {\bf 78}, 329 (1961).

12. H. Ray and A. S. Ghosh, J. Phys. B, {\bf 29}, 5505 (1996).

13. H. Ray and A. S. Ghosh, J. Phys. B, {\bf 30}, 3745 (1997).

14. A. S. Ghosh, N. C. Sil and P. Mandal, Phys. Rep., {\bf 87}, 313 (1982).

15. H. Ray and Ghosh A.S., J. Phys. B, {\bf 31}, 4427 (1998).
 
16. H. Ray, J. Phys. B, {\bf 32}, 5681 (1999).
 
17. H. Ray, J. Phys. B,{\bf 33}, 4285 (2000).

18. H. Ray, J. Phys. B, {\bf 35}, 2625 (2002).

19. Triplet resonance in Ps and H Scattering, Hasi Ray, Book of abstracts
XXIV ICPEAC held at Rosario, Argentina on July 20-26, 2005.   

20. Resonances in Ps and H scattering, Hasi Ray, Book of abstracts XIII International Workshop on Low-energy Positron and Positronium Physics, held at Campinas, SP, Brasil on July 27-30, 2005. 

21. Resonances in Ps-H Scattering, Hasi Ray, Book of abstracts EGAS37, held at Dublin, Ireland on August 3-6, 2005.  

22. New findings in Ps-H scattering, Hasi Ray, Book of abstracts ECAMP8, held at Rennes, France on July 6-10, 2004 p. 3-112. 

23. Resonances on Projectile-inelastic CCA in Ps-H scattering, Hasi Ray, Book of abstracts XXIII ICPEAC, held at Stockholm University, Sweden on July 23-29, 2003 p. Th. 104. 

24. New findings in Ps-H scattering using exact CCA theory, H. Ray, Phys. Rev. Lett. Manuscript No. LK9391 (2003-2004) unpublished. 

25. A new kind of binding, H. Ray, ISAMP NEWS LETTER, Vol.I Issue 2 p.6-8 (2005). 

26. P. K. Sinha, P. P. Chaudhury and A. S. Ghosh, J. Phys. B, {\bf 30}, 4644 (1997).

27. A. S. Ghosh, P. K. Sinha and H. Ray, Nucl. Inst. Meth. B, {\bf 143}, 162 (1998).

28. S. K. Adhikari and P. K. Biswas, Phys. Rev. A, {\bf 59}, 2058 (1999).

29. P. A. Fraser, Proc. Phys. Soc., {\bf 79}, 721 (1962).

30. P. A. Fraser, J.Phys. B, {\bf 1} 1006 (1968). 

31. P. A. Fraser and M. Kraidy, Proc. Phys. Soc., {\bf 89}, 533 (1966).

32. S. Hara and P. A. Fraser, J. Phys. B, {\bf 8}, L472 (1975).

33. John Maddox, Nature, {\bf 314}, 315 (1985).

34. B. H. Bransden, Case Studies ed E. W. McDaniel and M. R. C. McDowell (Amsterdam: North-Holland), {\bf 1}, 169-248 (1969).

35. P. G. Burke, H. M. Schey and K. Smith, Phys. Rev., {\bf 129}, 1258 (1963).

36. P. G. Burke and A. J. Taylor, J.Phys.B, {\bf 2}, 869 (1969).

37. A. N. Tripathy, K. C. Mathur and S. K. Joshi, Phys. Rev. A, {\bf 4}, 1873 (1971).

38. H. R. J. Walters, J. Phys. B, {\bf 9}, 227 (1976). 

39. R. J. Drachman and A. K. Bhatia, Phys. Rev. A, {\bf 51}, 2926 (1995).

40. A.S. Ghosh and G. Sunanda, Phys. Rev. A, {\bf 28}, 743 (1981).

41. N.F.Mott and H.S.W.Massey, The Theory of Atomic Collisions, Third Edition (Reprinted 1987) Vol.II p.414. 

42. H. Ray, Phys. Lett. A {\bf 252}, 316 (1999).

43. H. Ray, Phys. Lett. A {\bf 299}, 65 (2002).

44. H. Ray, Nucl. Inst. Meth. B, {\bf 192}, 191 (2002).

45. H. Ray, J. Phys. B, {\bf 35}, 3365 (2002). 

46. H. Ray, PRAMANA, {\bf 63}, 1063 (2004). 

47. H. Ray, PRAMANA, {\bf 66}, 415 (2006).

48. H. Ray, Euro. Phys. Lett., {\bf 73}, 21 (2006). 

49. C. P. Campbell, M. T. McAlinden, F. R. G. S. MacDonald and H. R. J. Walters, Phys. Rev. Lett., {\bf 80}, 5097 (1998).

50. R. J. Drachman and S. K. Houston, Phys. Rev. A, {\bf 14}, 894 (1976).

51. J. E. Blackwood, M. T. McAlinden and H. R. J. Walters, 
Phys. Rev. A, {\bf 65}, 32517 (2002).
 
52. B. A. P. Page, J. Phys. B, {\bf 9}, 1111 (1976).

53. R. J. Drachman, Phys. Rev. A, {\bf 19}, 1900 (1979).

54. J. DiRienzi and R. J. Drachman, Phys. Rev. A, {\bf 65}, 032721 (2002). 

55. J. DiRienzi and R. J. Drachman, Phys. Rev. A, {\bf 66}, 054702 (2002).

56. Z. C. Yanand Y. K. Ho, Phys. Rev. A, {\bf 59}, 2697 (1999).

57. Z. C. Yan and Y. K. Ho, Phys. Rev. A, {\bf 60}, 5098 (1999).

58. D. M. Schrader, F. N. Jacobsen, N. P. Frandsen and U. Mikkelsen, Phys. Rev. Lett., {\bf 69}, 57 (1992).

59. R. J. Drachman, private communication (2006).

60. G. G. Ryzhikh and J. Mitroy, Phys. Rev. Lett., {\bf 79}, 4124 (1997).

61. J. Mitroy, M. W. J. Bromley and G. G. Ryzhikh, Topical Review, J.Phys.B, {\bf 35}, R81 (2002).

62. I. Ivanov, J. Mitroy and K. Varga, Phys. Rev. A, {\bf 65}, 032703 (2002).

63. V. A. Dzuba, V. V. Flambaum, G. F. Gribakin and W. A. King, Phys. Rev. A, {\bf 52}, 4541 (1995).

64. P. V. Reeth and J. W. Humberston, J.Phys.B {\bf 36}, 1923 (2003).

65. S. Chiesa, M. Mella and G. Morosi, Phys. Rev. A, {\bf 66}, 042502 (2002).

\end{document}